\documentclass[conference]{IEEEtran}
\IEEEoverridecommandlockouts

\usepackage{cite}
\usepackage{amsmath,amssymb,amsfonts}
\usepackage{algorithmic}
\usepackage{graphicx}
\usepackage{textcomp}
\usepackage{xcolor}

\usepackage{amsthm}
\usepackage{algorithm}
\def\BibTeX{{\rm B\kern-.05em{\sc i\kern-.025em b}\kern-.08em
    T\kern-.1667em\lower.7ex\hbox{E}\kern-.125emX}}

\newcommand \0 {\mathbf 0}

\newcommand\blkdg{\ensuremath{{\rm blkdiag}}}

\newcommand\st{\ensuremath{{\rm s.t.}}}
\def\Vec{\mathrm{vec}}

\newtheorem{Lemma}{Lemma}

\newcommand\Ch{\ensuremath{\mathbb{C}}}

\newcommand\Cc{\ensuremath{\mathcal{C}}}
\newcommand\Fc{\ensuremath{\mathcal{F}}}

\newcommand\Nc{\ensuremath{\mathcal{N}}}

\def\a {\mathbf a}	

\def\C {\mathbf C}

\def\F {\mathbf F}

\def\H {\mathbf H}	
\def\h {\mathbf h}	

\def\I {\mathbf I}

\def\K {\mathbf K}

\def\s {\mathbf s}

\def\U {\mathbf U}
\def\u {\mathbf u}	%

\def\X {\mathbf X}	  
\def\x {\mathbf x}    

\def\Y {\mathbf Y}
\def\y {\mathbf y}

\def\Z {\mathbf Z} 
\def\z {\mathbf z}

\def\bomega {\boldsymbol \omega}

\def\blambda {\boldsymbol \lambda}
\def\hlambda {\hat{\lambda}}
\newcommand\bLambda{\ensuremath{{\bf \Lambda}}}

\begin{document}

\title{Symbol-Level Precoding-Based Self-Interference Cancellation for ISAC Systems 
}

\author{\IEEEauthorblockN{Shu Cai\IEEEauthorrefmark{1}, Zihao Chen\IEEEauthorrefmark{1}, Ya-Feng Liu\IEEEauthorrefmark{2}, and Jun Zhang\IEEEauthorrefmark{1}}
\IEEEauthorblockA{\IEEEauthorrefmark{1}Jiangsu Key Lab of Wireless Communications, Nanjing University of Posts and Telecommunications, Nanjing, China\\
Engineering Research Center of Health Service System Based on Ubiquitous Wireless Networks, Nanjing, China\\
Email: \{caishu, 1222013939, zhangjun\}@njupt.edu.cn}
\IEEEauthorblockA{\IEEEauthorrefmark{2}LSEC, ICMSEC, AMSS, Chinese Academy of Sciences, Beijing, China\\
Email: yafliu@lsec.cc.ac.cn}
}

\maketitle

\begin{abstract}
Consider an integrated sensing and communication (ISAC) system where a base station (BS) employs a full-duplex radio to simultaneously serve multiple users and detect a target. The detection performance of the BS may be compromised by self-interference (SI) leakage. This paper investigates the feasibility of SI cancellation (SIC) through the application of symbol-level precoding (SLP). We first derive the target detection probability in the presence of the SI. 
We then formulate an SLP-based SIC problem, which optimizes the target detection probability while satisfying the quality of service requirements of all users. 
The formulated problem is a nonconvex fractional programming (FP) problem with a large number of equality and inequality constraints. We propose a penalty-based block coordinate descent (BCD) algorithm for solving the formulated problem, which allows for efficient closed-form updates of each block of variables at each iteration. 
Finally, numerical simulation results are presented to showcase the enhanced detection performance of the proposed SIC approach.
\end{abstract}


\section{Introduction}

Integrated sensing and communication (ISAC) is a pivotal technology for advancing 5G-Advanced and 6G networks \cite{Liu2022JSAC}. Its significance lies in its ability to enhance both spectral and energy efficiency by sharing hardware and resources between sensing and communication (S\&C) functions \cite{Liu2020TC}. 
Transmit beamforming (TBF) is a key signal processing technique to achieve these efficiency improvements and has been widely researched \cite{Fan2018TSP,Chen2022,Liu2022TSP,Fan2020WCL,LiMing2022JSAC}. 
The work \cite{Fan2018TSP} focused on transmit waveform design in downlink ISAC systems and investigated the S\&C performance tradeoff. 
TBF designs have been extended to ISAC systems with more practical sensing performance metrics, such as the sensing signal-to-interference-plus-noise ratio (SINR) \cite{Chen2022}, Cramér-Rao bound \cite{Liu2022TSP}, and range sidelobe \cite{Fan2020WCL}, to name a few.
The work \cite{LiMing2022JSAC} further considered transceiver design, which optimizes the symbol-level precoding (SLP) and receiver filter to maximize the sensing SINR. 
In the referenced ISAC systems, the base station (BS) receives target echoes while transmitting signals, i.e., it operates in full-duplex (FD) mode \cite{Andrew2022CST}. 
In this configuration, the transmission signal can mix with the received signal, which is typically much stronger than the target echoes, thereby leading to significant self interference (SI).
 
A common assumption in most of existing works on FD ISAC systems is that the SI is adequately mitigated through techniques like natural isolation, analog cancellation, and digital processing \cite{smida_-band_2024,Baquero2019TMTT}. However, the residual SI can still overpower desired signals \cite{smida_-band_2024}, which poses a significant challenge to the efficacy of ISAC systems.
Recent works have focused on beamforming techniques in FD ISAC systems to mitigate the SI \cite{he_full-duplex_2023,wang_bidirectional_2024}. 
In particular, the work \cite{he_full-duplex_2023} proposed to integrate FD capabilities into both S\&C, optimizing the downlink dual-functional transmit signal and uplink receive beamformers at the BS, along with the transmit power at uplink users. The work \cite{wang_bidirectional_2024} examined TBF in bidirectional ISAC systems across narrowband and wideband scenarios, comparing the performance of full-duplex and half-duplex operations.

In this paper, we consider an ISAC system with one FD BS serving multiple downlink users and at the same time detecting a target, and investigate the SLP design for SI cancellation (SIC). 
In sharp contrast to previous works \cite{he_full-duplex_2023,wang_bidirectional_2024}, which employ the SINR as the radar performance metric, we first derive the target detection probability in the presence of the SI and then formulate an SLP-based SIC problem, which optimizes the derived detection probability
subject to all users’ quality of service requirement constraints. 
The formulated problem is challenging to solve, as it is a nonconvex fractional program (FP) with numerous equality and inequality constraints. By carefully exploiting the problem's special structures, we develop a penalty-based block coordinate descent (BCD) algorithm, which yields closed-form optimization at each iteration. 
Numerical results demonstrate that the proposed method brings tremendous improvements in terms of the detection probability.

\section{System and Signal Model}
\begin{figure}[!t]
	\begin{center}
		{
			{\includegraphics[width=0.26\textwidth]{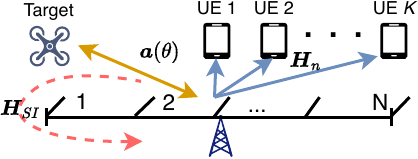}}}
	\end{center}\vspace{-0.5cm}
	\caption{A MIMO-OFDM ISAC system.}
	\vspace{-0.3cm}
	\label{p1}
\end{figure}

Consider the downlink MIMO-OFDM system as depicted in Fig. \ref{p1}. The BS is adorned with a uniform linear array (ULA) consisting of $M$ antennas, enabling it to concurrently serve $K$  single-antenna communication users and detect a target. 
\subsection{Communication Signal Model}
The BS is designed to transmit a sequence of $N$ QPSK\footnote{It is straightforward to extend the results of this paper from the QPSK case to the general PSK case.} symbols, denoted as $s_{k,1},s_{k,2},\dots,s_{k,N}$ for each user $k$, using OFDM signals with $N$ subcarriers. Denote the data on the $n$-th subcarrier as $\s_n=[s_{1,n},s_{2,n},\dots,s_{K,n}]^T$. It is firstly precoded to generate the SLP vector $\x[n]$. Then, let $\X = [\x[1],\x[2],\dots,\x[N]]\in\Ch^{M\times N}$, and feed this sequence to each antenna for OFDM modulation to generate the time-domain signals \cite{Yuanyuan2024}: 
    $\U \triangleq [\u[1],\u[2],\dots,\u[N]] = \X\F$,
where $\F\in \Ch^{N \times N}$ denotes the conjugate of the fast Fourier transform (FFT) matrix with $\F\F^H = \I$. 

Consider the frequency-selective fading effects, and let 
the MIMO channel on the $n$-th subcarrier be characterized by the matrix $\H_n \in \Ch^{K \times M}$, where $n = 1,2,\dots,N$. 
Therefore, the received signals for all $K$ users at the $n$-th subcarrier after OFDM demodulation can be represented by 
\par\vspace{-1ex}
{\small \begin{equation}\label{eq:comsig1}
\y_c[n]=\H_n\x[n]+\z_c[n],
\end{equation}}%
where $\z_c[n] \sim \Cc\Nc\left(\0, \sigma_c^{2} \I\right)$ denotes the additive white Gaussian noise (AWGN).

Assume that the channel state information is known to the BS. Then, the SLP designs $\x[n]$ to ensure that the output after passing through the MIMO channel satisfies \cite{Ang2021TIFS,Cai2022TSP}
\par\vspace{-1ex}
{\small \begin{equation}\label{eq:TPcon1}
\H_n\x[n]=\blambda_n \odot \s_n,~\forall~n=1,2,\dots,N.  
\end{equation}}%
Here, $\blambda_n=[\lambda_{1,n},\lambda_{2,n},\dots,\lambda_{K,n}]^T$ and the complex scalar $\lambda_{k,n}$ act as slack variables, ensuring that the received signal $\lambda_{k,n} s_{k,n}$ falls within the construction region \cite{Ang2021TIFS} of the constellation point $s_{k,n}$. For QPSK symbols, this requires that $\lambda_{k,n}$ must satisfy the constraint 
\cite{Ang2021TIFS}
\par\vspace{-1ex}
{\small \begin{equation}\label{eq:TPcon2}
\big|\lambda_{k,n}^{\Im}\big|\le \tan(\phi)\big(\lambda_{k,n}^{\Re}-\Gamma_{k}\big),~\forall~ k,n,
\end{equation}}%
where $\phi = \frac{\pi}{4}$, $\lambda_{k,n}^{\Re}$ and $\lambda_{k,n}^{\Im}$ denote the real and imaginary parts of a complex number $\lambda_{k,n}$, respectively, and the distance from the thresholds for signal detection to the constructive region is determined as $\Gamma_{k} = \sqrt{\gamma_k \sigma_{C}^{2}}$, with $\gamma_k$ representing the SINR requirement of user $k$.

\subsection{Radar Signal Model}
Consider the scenario where the ISAC BS aims to detect a potential target located at distance $d_t$. The round-trip time delay of the signal can be obtained by $\tau_t=2d_t/v_c$, where $v_c$ is the velocity of light. Denoting the duration of each $\u[n]$ as $T_s$. Then the normalized time delay is $n_t \triangleq \lfloor \frac{\tau_t}{T_s} \rfloor$, and the output of the radar receiver, represented by
$\Y_s\triangleq\left[\y_s[1+n_t], \y_s[2+n_t], \dots, \y_s[N+n_t]\right]$, is \par\vspace{-2ex}
{\small \begin{align}\label{eq:radsig3}
\Y_s=\beta\a(\theta)\a^{T}(\theta)\X\F + \H_{SI}\X\F_{n_t} + \Z_s.
\end{align}}%
In \eqref{eq:radsig3}, $\beta\in \Ch$ denotes the attenuation factor including the path loss and the target radar cross section (RCS), $\H_{SI}$ denotes the residual SI channel after analog/digital SIC \cite{ChangTWC2019}, $\F_{n_t}=\F\I_{n_t}$, $\I_{n_t}$ is a $N$-dimensional matrix with ones on the $n_t$-th diagonal below the main diagonal and zeros elsewhere, the entries of the noise matrix $\Z_s$ are Gaussian distributed with $\Cc\Nc(0,\sigma_{s}^{2})$,  and 
 $\a(\theta)= [1,e^{-j\frac{2\pi d_a}{\lambda} \sin(\theta)},\dots,e^{-j\frac{2\pi d_a}{\lambda}(M-1) \sin(\theta)}]^T$
denotes the steering vector from the ULA to the target, where $\lambda$ denotes the wavelength, $d_a$ is the antenna spacing, and $\theta$ denotes the direction of arrival (DoA) of the target echo. For simplicity, we assume that $K \le M$, $d_a = 0.5\lambda$, and the entries of $\H_{SI}$ are identically and independently distributed (i.i.d.) and Gaussian distributed with $\Cc\Nc(0,\sigma_{SI}^{2})$.  This i.i.d. Gaussian distribution assumption on the residual SI channel is crucial in simplifying our following derivation. In practice, this assumption might not always hold true. In this case, we can adopt a robust optimization technique as in \cite{Tsung-Hui2017SPAWC} to deal with the uncertainty on the residual SI channel. We shall explore this in our future work.

\section{Problem Formulation}

\subsection{Target Detection with the SI} 
Target detection based on \eqref{eq:radsig3} is difficult, since the SI is correlated both in space and time. Hence, a prewhitening process is required.
Denote the vectorization of $\Y_s$ as $\y_s=\Vec(\Y_s)$. Then 
\par\vspace{-2ex}
{\small \begin{align}\label{eq:radsig4}
\y_s = \beta((\X\F)^T\otimes \I) \a_E(\theta)+\z,
\end{align}}%
where $\a_E(\theta) = \a(\theta)\otimes\a(\theta)$, 
$\z=((\X\F_{n_t})^T\otimes \I)\h_{SI}+\z_s$, 
$\h_{SI} = \Vec(\H_{SI})$, $\z_s=\Vec(\Z_s)$, and $\I$ is the identity matrix.  
Since both $\h_{SI}$ and $\z_s$ are i.i.d. Gaussian variables, it follows that $\z$ is also Gaussian distributed with mean zero and variance
\par\vspace{-2ex}
{\small \begin{align}\label{eq:radSINoiVar}
	\C_{n_t} &=  \K^T \otimes \I,
\end{align}}%
where $\K=\sigma_{SI}^{2}\F_{n_t}^H\X^H\X\F_{n_t}+\sigma_s^2 \I$.
By letting $\y_w = \C_{n_t}^{-\frac{1}{2}}\y_s$, we can obtain $\z_w=\C_{n_t}^{-\frac{1}{2}}\z\sim\Cc\Nc(\0,\I)$.
Based on $\y_w$, the signals for the presence ($H_1$) and absence ($H_0$) of the target at the distance $d_t$ are modeled as
\par\vspace{-2ex}
{\small \begin{align}\label{eq:whitesig2}
	\y_w = \left\{
		\begin{aligned}
			&\beta\C_{n_t}^{-\frac{1}{2}}((\X\F)^T\otimes \I) \a_E(\theta)+\z_w, &(H_1)\\
			&\z_w, &(H_0)
		\end{aligned}
		\right..
\end{align}}%
Target detection can then be performed using the generalized likelihood ratio test (GLRT)\footnote{The BS aims to detect a target at a known and fixed direction $\theta$, with the exact DoA unknown. Therefore, the GLRT is employed for estimation and detection.} \cite{kay1998fundamentals}, as described in Lemma \ref{lemma:LLR}.
\begin{Lemma}\label{lemma:LLR}
    Denote $\theta^{\star}$ as the optimal solution of
    \par\vspace{-2ex}
    {\small \begin{align}\label{eq:LLR2}
	\max_{\theta}~\left\{L_{n_t}(\theta) \triangleq \frac{|\a^H(\theta)\Y_s\K^{-1}\F^H\X^H\a^*(\theta)|^2}{M\a^T(\theta)\X\F\K^{-1}\F^H\X^H\a^*(\theta)}\right\},
    \end{align}}%
    and $\a = \a(\theta^{\star})$. 
    Then, the GRLT based on \eqref{eq:whitesig2} is given by  
 \par\vspace{-2ex}
    {\small \begin{align}\label{eq:LLR2add}
	L_{n_t}(\theta^{\star})~\underset{H_1}{\overset{H_0}{\lessgtr}} \delta~ \textrm{with}~ 
 L_{n_t}(\theta^{\star})  \sim  \left\{
		\begin{aligned}
			&\chi^2_2(\rho) , &(H_1)\\
			&\chi^2_2, &(H_0)
		\end{aligned}
		\right..
\end{align}}%
Here, $\delta$ is the threshold for target detection,  
 $\chi^2_2$ and $\chi^2_2(\rho)$ denote the central and non-central chi-squared distributions with two degrees of freedom, respectively, and 
	$\rho = M|\beta|^2\a^T\X\F\K^{-1}\F^H\X^H\a^*$.
\end{Lemma}
\begin{proof}
    The proof is based on results and techniques in \cite{Khawar2015,kay1998fundamentals}. We omit the detailed proof due to the space reason.
\end{proof}

According to Lemma \ref{lemma:LLR}, the detection probability under a desired false alarm probability, $P_{FA}$, can be determined using the Neyman-Pearson criterion \cite{Khawar2015}, which is expressed as
\par\vspace{-2ex}
 {\small \begin{align}\label{eq:PD}
	P_D = 1 - \Fc_{\chi_2^2(\rho)}(\delta),
\end{align}}%
where $\delta = \Fc_{\chi_2^2}^{-1}(1-P_{FA})$, $\Fc_{\chi_2^2}^{-1}$ is the inverse central chi-squared distribution function, and $\Fc_{\chi_2^2(\rho)}$ is the noncentral chi-squared distribution function.
\vspace{-0.5ex}
\subsection{SLP-based SIC Problem Formulation}
According to  \eqref{eq:PD}, when $P_{FA}$ is given, $P_D$ increases as $\rho$ increases. Therefore, 
we can formulate the SLP problem of maximizing the detection probability as follows: 
\par\vspace{-2ex}
{\small \begin{subequations}\label{eq:TP}
	\begin{align}
		\max_{\X,\bLambda}~  & \a^T\X\F\K^{-1}\F^H\X^H\a^* \label{eq:TPa}\\
	\st ~      & \|\X\|_F^2\leq P_T,  \label{eq:TPb}\\
    &\H_n\x[n]=\blambda_n \odot \s_n,~\forall~ n=1,2,\dots,N,  \label{eq:TPc}\\
	&\left|\lambda_{k,n}^{\Im}\right|\le \tan (\phi)\left(\lambda_{k,n}^{\Re}-\Gamma_{k}\right),~\forall~ k,n, \label{eq:TPd}
	\end{align}
\end{subequations}}%
where $P_T$ is the power budget, \eqref{eq:TPc} and \eqref{eq:TPd} are constraints for guaranteeing the communication performance \cite{Ang2021TIFS}. 
The formulated SLP-based SIC problem \eqref{eq:TP} is challenging to solve, as it is nonconvex (e.g., there is an FP term in the objective) and has a large number of equality and inequality constraints. 

\section{The Proposed Approach}
In this section, we propose an efficient algorithm for solving problem \eqref{eq:TP}, which leverages the penalty method \cite{liu2024survey} and the block coordinate descent method. The idea is to penalize all equality constraints in \eqref{eq:TPc} to the objective via the quadratic penalty function and solves problem \eqref{eq:TP} by solving a sequence of (relatively easy) penalty problems as follows: 
\par\vspace{-2.5ex}
{\small \begin{subequations}\label{eq:TPMM1}
	\begin{align}
		\max_{\X,\bLambda}~& \a^T\X\F\K^{-1}\X^H\F^H\a^* - \varrho \sum_{n=1}^N\|\H_n\x[n]-\blambda_n \odot \s_n\|^2 \label{eq:TPMM1a}\\
	\st ~ & \eqref{eq:TPb},~\eqref{eq:TPd},
	\end{align}
\end{subequations}}%
where $\varrho>0$ is the penalty parameter. The penalty algorithm for solving problem \eqref{eq:TP} is presented in Algorithm \ref{alg:TPPM}.
\begin{algorithm}[!t]
\caption{The Penalty Method for Problem \eqref{eq:TP}.} 
\begin{algorithmic}[1]\label{alg:TPPM}
\STATE Initialize $i=1$, $c_{\varrho}>1$, and $\varrho$ and $\epsilon_p$ as small numbers.
\REPEAT
    \STATE Solve problem \eqref{eq:TPMM1} to obtain $\X^{(i)}$ and $\bLambda^{(i)}$.
    \STATE Let $\varrho = c_{\varrho}\varrho$ and $i = i+1$,
    \UNTIL some stopping criterion is satisfied.
\end{algorithmic}
\end{algorithm}

The subsequent part of this section focuses on solving problem \eqref{eq:TPMM1}. We first use the quadratic transform \cite{liu2024survey,KaimingFP2} to overcome the difficulty of the FP term in the objective function of problem \eqref{eq:TPMM1} and obtain the following equivalent form:
\par\vspace{-2ex}
{\small \begin{subequations}\label{eq:SICPd2}
	\begin{align}
		\min_{\y,\X,\bLambda}~  & \varrho\sum_{n=1}^N\|\H_n\x[n]-\blambda_n \odot \s_n\|^2 - 2\Re\{\y^H\F^H\X^H\a^*\} \notag\\ 
          &\qquad + \y^H(\sigma_{SI}^{2}\F_{n_t}^H\X^H\X\F_{n_t}+\sigma_R^2 \I)\y \label{eq:SICPd2a}\\
	\st\quad  
     &\eqref{eq:TPb},~\eqref{eq:TPd}.
	\end{align}
\end{subequations}}%
A key structural feature of problem \eqref{eq:SICPd2} is that the three blocks of variables in it are well decoupled and the update of each block of variables (with the other two being fixed) admits efficient (semi-)closed-form solutions. Next, we will present the closed-form solutions of each update of variables one by one.
\vspace{-0.5ex}
\subsection{Closed-Form Solution of the $\y$-Subproblem}
When $\X$ and $\bLambda$ are given, the subproblem of updating $\y$ has a closed-form solution, which is
\par\vspace{-2ex}
{\small \begin{align}\label{eq:opty}
    \y = (\sigma_{SI}^{2}\F_{n_t}^H\X^H\X\F_{n_t}+\sigma_R^2 \I)^{-1}\F^H\X^H\a^*.
\end{align}}%

\subsection{Closed-Form Solution of the $\X$-Subproblem }\label{subsec:TPCUpdateX}
Denote $\y_{n_t} = \sigma_{SI}\F_{n_t}\y$. Then the subproblem of updating $\X$ is 
\par\vspace{-2ex}
{\small \begin{subequations}\label{eq:SICX}
	\begin{align}
		\min_{\X}~  & \varrho \sum_{n=1}^N\|\H_n\x[n]-\blambda_n \odot \s_n\|^2 + \y^H_{n_t}\X^H\X\y_{n_t} \notag\\ 
          &~~~- 2\Re\{\y^H\F^H\X^H\a^*\}  \label{eq:SICXa}\\
	\st ~ & \|\X\|_F^2\leq P_T,  \label{eq:SICXb}
	\end{align}
\end{subequations}}%
which is a convex quadratic program with a quadratic inequality constraint. The Lagrangian of problem \eqref{eq:SICX} is 
\par\vspace{-2ex}
{\small \begin{align}\label{eq:Lag}
f(\X,\eta) =\;& \varrho\sum_{n=1}^N\|\H_n\x[n]-\blambda_n \odot \s_n\|^2 + \y_{n_t}^H\X^H\X\y_{n_t}  \notag\\ 
&~~~ - 2\Re\{\y^H\F^H\X^H\a^*\} + \eta\|\X\|_F^2-\eta P_T, 
\end{align}}%
where $\eta\geq 0$ is the Lagrange multiplier corresponding to the inequality constraint \eqref{eq:SICXb}. 
By setting the first-order derivative of \eqref{eq:Lag} to be zero, i.e., $\nabla_\X f(\X,\eta) = 0$, and leveraging the Kronecker product property \cite{GoluVanl96}, the optimal $\X$ can be obtained as follows:
\par\vspace{-2ex}
{\small \begin{align}\label{eq:SICX4S1}
\x(\eta) = [(\y^*_{n_t}\y^T_{n_t})\otimes \I +  \varrho\H^H\H+\eta\I]^{-1}\bomega,
\end{align}}%
where $\x = \Vec(\X)$, $\H=\blkdg(\H_1,\H_2,\dots,\H_N)$, $\bomega = \Vec(\a^*\y^H\F^H+\varrho\H^H(\blambda \odot \s))$, $\s = \Vec([\s_1,\s_2,\dots,\s_N])$, $\blambda = \Vec([\blambda_1,\blambda_2,\dots,\blambda_N])$, 
and the desired $\eta$ satisfies 
$\eta = 0$, if $\|\x(0)\|_F^2\leq P_T$; otherwise, $\eta=\eta^{\star}$ with $\|\x(\eta^{\star})\|_F^2= P_T$.
Since $\|\x(\eta)\|^2$ is a monotocially decreasing function of $\eta$ according to \eqref{eq:SICX4S1}, the optimal $\eta^{\star}$ can be efficiently obtained by bisection. 
\vspace{-0.5ex}
\subsection{Closed-Form Solution of the $\blambda$-Subproblem}
In cases of PSK constellation, we have $\|\H_n\x[n]-\blambda_n\odot\s_n\| =  \|\hat{\blambda}_n - \blambda_n \|$, where $\hat{\blambda}_n = \H_n\x[n]\odot \s_n^*$. As a result,
the subproblem of updating $\blambda_n$ can be written as 
\par\vspace{-2ex}
{\small \begin{subequations}\label{eq:TPMM2PLa}
	\begin{align}
		\min_{\blambda_n}~  & \|\hat{\blambda}_n - \blambda_n \|^2 \label{eq:TPMM2PLaa}\\
	\st ~ &\left|\lambda_{k,n}^{\Im}\right|\le \tan (\phi)\left(\lambda_{k,n}^{\Re}-\Gamma_{k}\right),~\forall~ k. 
	\end{align}
\end{subequations}}%
Then problem \eqref{eq:TPMM2PLa} is decoupled into $K$ subproblems with the following form:
{\small \begin{subequations}\label{eq:TPMM2PLa2}
\begin{align}
		\min_{\lambda}~  & |\hat{\lambda} - \lambda| \label{eq:TPMM2PLa2a}\\
	\st ~ &\left|\lambda^{\Im}\right|\le \tan (\phi)\left(\lambda^{\Re}-t\right).\label{eq:TPMM2PLa2b}
\end{align}
\end{subequations}}%
The optimal solution of \eqref{eq:TPMM2PLa2} is given in the following lemma.

\begin{Lemma}\label{lemma:optla}
	Denote the optimal solution of \eqref{eq:TPMM2PLa2} as $\lambda_*$. Then we have the following four cases: 
     \begin{itemize}
\item[$\bullet$] if $\frac{\hlambda^{\Re}-t}{|\hlambda^{\Im}|} \ge \cot(\phi)$, then $\lambda_*=\hlambda$;
\item[$\bullet$] if $\cot{\left(\phi+\frac{\pi}{2}\right)}<\frac{\hlambda^{\Re}-t}{\hlambda^{\Im}}<\cot(\phi)$ and $\hat\lambda^{\Im}>0$, then 
		{\small \begin{align}\label{eq:caseB}
    &\lambda_*^{\Re}=\hat\lambda^{\Re}\cos^2(\phi)+\hat\lambda^{\Im}\cos(\phi)\sin(\phi)+t\sin^2(\phi),\\
		&\lambda_*^{\Im}=(\hat\lambda^{\Re}-t)\cos(\phi)\sin(\phi)+\hat\lambda^{\Im}\sin^2(\phi);
		\end{align}}%
\item[$\bullet$] if $\cot{\left(\phi+\frac{\pi}{2}\right)}<\frac{\hlambda^{\Re}-t}{-\hlambda^{\Im}}<\cot(\phi)$ and $\hlambda^{\Im}<0$, then 
	{\small\begin{align}\label{eq:caseC}
	&\lambda_{*}^{\Re}=\hat\lambda^{\Re}\cos^2(\phi)-\hat\lambda^{\Im}\cos(\phi)\sin(\phi)+t\sin^2(\phi)\\
	&\lambda_{*}^{\Im}=(t-\hat\lambda^{\Re})\cos(\phi)\sin(\phi)+\hat\lambda^{\Im}\sin^2(\phi);
	\end{align}}%
    \item[$\bullet$] if $\frac{\hlambda^{\Re}-t}{|\hlambda^{\Im}|} \le \cot{\left(\phi+\frac{\pi}{2}\right)}$, then $\lambda_{*}=t$.
\end{itemize}
\end{Lemma}
\begin{proof}
Please refer to Section IV-B of \cite{HouICCC2022}.
\end{proof}

Finally, we summarize the BCD algorithm in Algorithm \ref{alg:TPBCD}.
\begin{algorithm}[!t]
\caption{The BCD Algorithm for Problem \eqref{eq:SICPd2}.} 
\begin{algorithmic}[1]\label{alg:TPBCD}
\STATE Initialize $i=0$, $\X^{(i)}$, and $\bLambda^{(i)}$. 
\REPEAT
    \STATE Let $i=i+1$,
    \STATE Update $\y^{(i)}$ by \eqref{eq:opty}, 
    \STATE Update $\X^{(i)}$ by \eqref{eq:SICX4S1},
    \STATE Update $\bLambda^{(i)}$ by using Lemma \ref{lemma:optla},
    \UNTIL some stopping criterion is satisfied.
\end{algorithmic}
\end{algorithm}

\section{Simulation Results} \label{sec: simulations}
In this section, we evaluate the performance of the proposed SLP design for SIC through simulations. We consider a BS equipped with $M=32$ antennas, transmitting QPSK signals to $K=8$ users. The BS operates with a power budget of $30$dBm, and the AWGN variances are set at $-90$dBm. The communication channels are assumed to undergo Rayleigh fading, with large-scale fading modeled as $140.7 + 36.7 \log_{10}(d)$dB, where $d$(km) represents the distance between the BS and the corresponding user \cite{ChangTWC2019}. The users are randomly and uniformly distributed within a circular area centered at the BS, with a radius of $100$m. 
The number of subcarriers is $N = 2^8$, each subcarrier has a bandwidth of $\Delta = 120$kHz, and the SINR requirement for each user is $\gamma_k = 10$dB. For target detection purposes, the duration of a symbol is $T_s = \frac{1}{N\Delta}$s, and the residual SI channel power is $\sigma_{SI}^{2}=-80$dBm. The DoA of the target is uniformly distributed between $29^{\circ}$ and $31^{\circ}$. The attenuation factor is modeled as $\beta = (C_0 d_t^{-\alpha})^2$, where $C_0 =20$dB denotes the path loss at a reference distance of $1$m, $\alpha = 2$ is the path-loss exponent, and the squared term accounts for the round-trip path loss.

We consider two benchmark schemes: (1) SLP without SI (WoSI); (2) SLP with SI (WiSI), but without SIC. We adopt the detection probability and root mean square error (RMSE) of DoA estimation as the performance metrics. All the results were obtained by averaging over $100$ independent channel realizations. 

\begin{figure}[tb]
	\begin{center}
			{\includegraphics[width=0.4\textwidth]{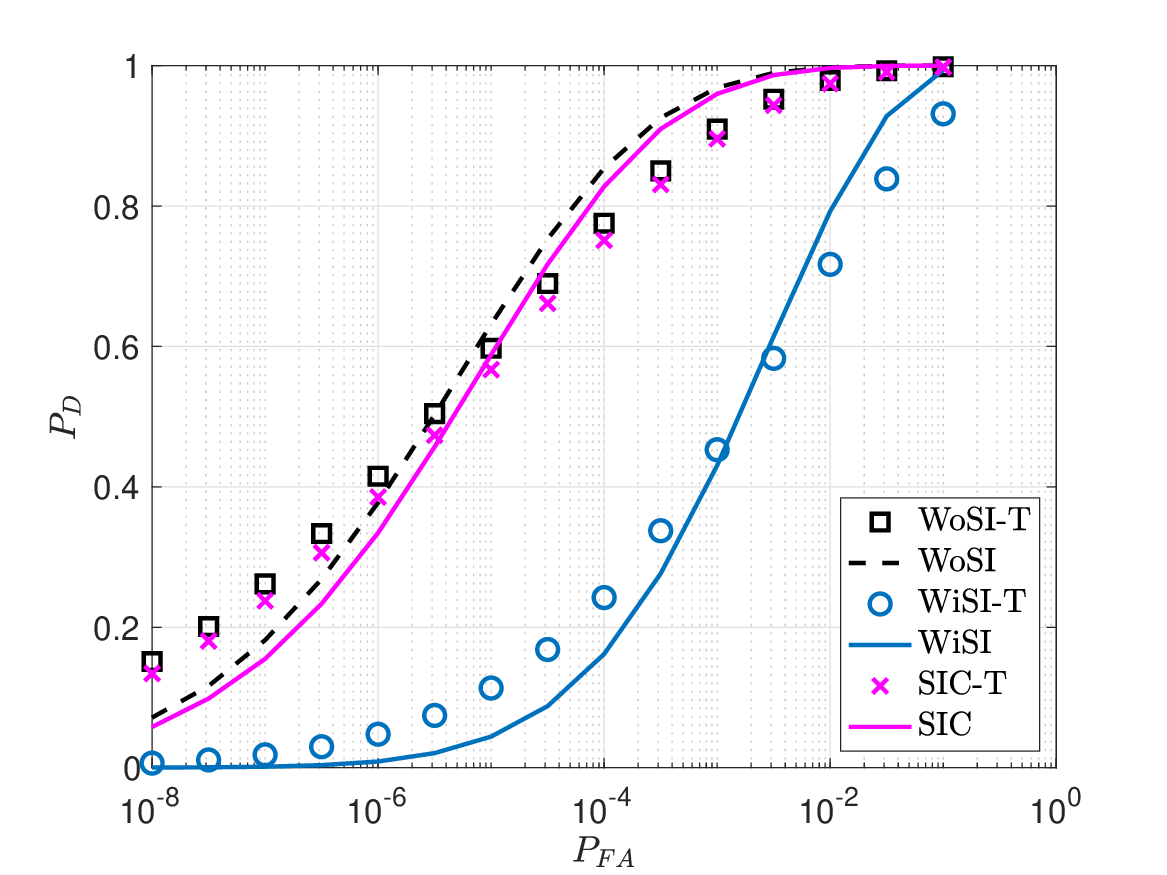}}
	\end{center}\vspace{-0.5cm}
	\caption{ROC of different methods.}\label{fig:TPRoC}
	\vspace{-0.3cm}
\end{figure}
%
\begin{figure}[tb]
	\begin{center}
			{\includegraphics[width=0.4\textwidth]{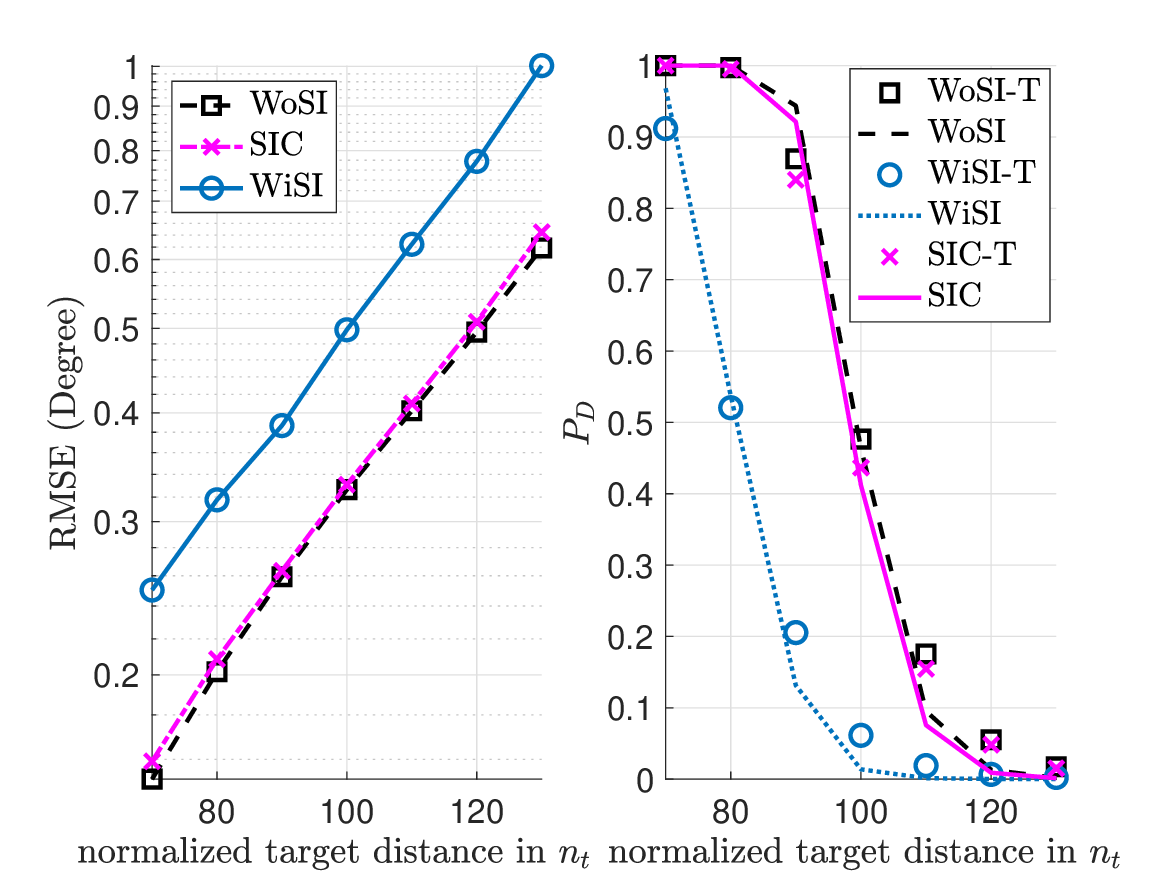}}
	\end{center}\vspace{-0.5cm}
	\caption{DoA estimation and detection accuracy versus the target distance.}\label{fig:DetTPTar}
	\vspace{-0.3cm}
\end{figure}

Fig. \ref{fig:TPRoC} plots the receiver operating characteristic (ROC) performance with the distance between the BS and the target being set as $10^3$m. The curves labeled with and without ``T'' represent the results predicted by the derived $P_D$ in \eqref{eq:PD} and the average of $10^4$ simulations of GLRT, respectively. It is clear that the theoretical and simulation results align closely with each other. The presence of the SI is shown to degrade the detection probability, but applying the SLP-based SIC technique effectively mitigates the degradation. 

Fig. \ref{fig:DetTPTar} depicts the estimation and detection performance as a function of the distance between the BS and the target. It can be observed from Fig. \ref{fig:DetTPTar} that the estimation and detection accuracy decreases with the target's distance from the BS. This is primarily due to the diminishing power of the echo signal. 
It can also observed from the figure that the proposed approach can actually achieve almost the same RMSE performance as in the case without the SI (i.e., the first benchmark), which showcases that our proposed approach can indeed effectively mitigate the SI.

\bibliographystyle{IEEEtran}
\bibliography{SC-SLP}

\begin{thebibliography}{10}
\providecommand{\url}[1]{#1}
\csname url@samestyle\endcsname
\providecommand{\newblock}{\relax}
\providecommand{\bibinfo}[2]{#2}
\providecommand{\BIBentrySTDinterwordspacing}{\spaceskip=0pt\relax}
\providecommand{\BIBentryALTinterwordstretchfactor}{4}
\providecommand{\BIBentryALTinterwordspacing}{\spaceskip=\fontdimen2\font plus
\BIBentryALTinterwordstretchfactor\fontdimen3\font minus
  \fontdimen4\font\relax}
\providecommand{\BIBforeignlanguage}[2]{{%
\expandafter\ifx\csname l@#1\endcsname\relax
\typeout{** WARNING: IEEEtran.bst: No hyphenation pattern has been}%
\typeout{** loaded for the language `#1'. Using the pattern for}%
\typeout{** the default language instead.}%
\else
\language=\csname l@#1\endcsname
\fi
#2}}
\providecommand{\BIBdecl}{\relax}
\BIBdecl

\bibitem{Liu2022JSAC}
F.~Liu, Y.~Cui, C.~Masouros, J.~Xu, T.~X. Han, Y.~C. Eldar, and S.~Buzzi,
  ``Integrated sensing and communications: Toward dual-functional wireless
  networks for {6G} and beyond,'' \emph{IEEE J. Sel. Areas Commun.}, vol.~40,
  no.~6, pp. 1728--1767, 2022.

\bibitem{Liu2020TC}
F.~Liu, C.~Masouros, A.~P. Petropulu, H.~Griffiths, and L.~Hanzo, ``Joint radar
  and communication design: Applications, state-of-the-art, and the road
  ahead,'' \emph{IEEE Trans. Commun.}, vol.~68, no.~6, pp. 3834--3862, 2020.

\bibitem{Fan2018TSP}
F.~Liu, L.~Zhou, C.~Masouros, A.~Li, W.~Luo, and A.~Petropulu, ``Toward
  dual-functional radar-communication systems: Optimal waveform design,''
  \emph{IEEE Trans. Signal Process.}, vol.~66, no.~16, pp. 4264--4279, 2018.

\bibitem{Chen2022}
L.~Chen, Z.~Wang, Y.~Du, Y.~Chen, and F.~R. Yu, ``Generalized transceiver
  beamforming for {DFRC} with {MIMO} radar and {MU-MIMO} communication,''
  \emph{IEEE J. Sel. Areas Commun.}, vol.~40, no.~6, pp. 1795--1808, 2022.

\bibitem{Liu2022TSP}
F.~Liu, Y.-F. Liu, A.~Li, C.~Masouros, and Y.~C. Eldar, ``$\text {Cramér-Rao}$
  bound optimization for joint radar-communication design,'' \emph{IEEE Trans.
  Signal Process.}, vol.~70, pp. 240--253, 2022.

\bibitem{Fan2020WCL}
F.~Liu, C.~Masouros, T.~Ratnarajah, and A.~Petropulu, ``On range sidelobe
  reduction for dual-functional radar-communication waveforms,'' \emph{IEEE
  Wireless Commun. Lett.}, vol.~9, no.~9, pp. 1572--1576, 2020.

\bibitem{LiMing2022JSAC}
R.~Liu, M.~Li, Q.~Liu, and A.~L. Swindlehurst, ``Joint waveform and filter
  designs for {STAP-SLP}-based {MIMO-DFRC} systems,'' \emph{IEEE J. Sel. Areas
  Commun.}, vol.~40, no.~6, pp. 1918--1931, 2022.

\bibitem{Andrew2022CST}
J.~A. Zhang, M.~L. Rahman, K.~Wu, X.~Huang, Y.~J. Guo, S.~Chen, and J.~Yuan,
  ``Enabling joint communication and radar sensing in mobile networks—{A}
  survey,'' \emph{IEEE Commun. Surveys Tuts.}, vol.~24, no.~1, pp. 306--345,
  2022.

\bibitem{smida_-band_2024}
B.~Smida, R.~Wichman, K.~E. Kolodziej, H.~A. Suraweera, T.~Riihonen, and
  A.~Sabharwal, ``In-band full-duplex: The physical layer,'' \emph{Proc. IEEE},
  vol. 112, no.~5, pp. 433--462, 2024.

\bibitem{Baquero2019TMTT}
C.~Baquero~Barneto, T.~Riihonen, M.~Turunen, L.~Anttila, M.~Fleischer,
  K.~Stadius, J.~Ryynänen, and M.~Valkama, ``Full-duplex {OFDM} radar with
  {LTE} and {5G} {NR} waveforms: Challenges, solutions, and measurements,''
  \emph{IEEE Trans. Microw. Theory Techn.}, vol.~67, no.~10, pp. 4042--4054,
  2019.

\bibitem{he_full-duplex_2023}
Z.~He, W.~Xu, H.~Shen, D.~W.~K. Ng, Y.~C. Eldar, and X.~You, ``Full-duplex
  communication for {ISAC}: Joint beamforming and power optimization,''
  \emph{IEEE J. Sel. Areas Commun.}, vol.~41, no.~9, pp. 2920--2936, 2023.

\bibitem{wang_bidirectional_2024}
Z.~Wang, X.~Mu, and Y.~Liu, ``Bidirectional integrated sensing and
  communication: Full-duplex or half-duplex?'' \emph{IEEE Trans. Wireless
  Commun.}, vol.~23, no.~8, pp. 8184--8199, 2024.

\bibitem{Yuanyuan2024}
Y.~Qin, A.~Li, Y.~Lyu, X.~Liao, and C.~Masouros, ``Symbol-level precoding for
  {PAPR} reduction in multi-user {MISO}-{OFDM} systems,'' \emph{IEEE Trans.
  Wireless Commun.}, 2024.

\bibitem{Ang2021TIFS}
Y.~Fan, A.~Li, X.~Liao, and V.~C.~M. Leung, ``Secure interference exploitation
  precoding in {MISO} wiretap channel: Destructive region redefinition with
  efficient solutions,'' \emph{IEEE Trans. Inf. Forensics Security}, vol.~16,
  pp. 402--417, 2021.

\bibitem{Cai2022TSP}
S.~Cai, H.~Zhu, C.~Shen, and T.-H. Chang, ``Joint symbol level precoding and
  receive beamforming optimization for multiuser {MIMO} downlink,'' \emph{IEEE
  Trans. Signal Process.}, vol.~70, pp. 6185--6199, 2022.

\bibitem{ChangTWC2019}
X.~Zhang, T.-H. Chang, Y.-F. Liu, C.~Shen, and G.~Zhu, ``Max-min fairness user
  scheduling and power allocation in full-duplex {OFDMA} systems,'' \emph{IEEE
  Trans. Wireless Commun.}, vol.~18, no.~6, pp. 3078--3092, 2019.

\bibitem{Tsung-Hui2017SPAWC}
T.-H. Chang, Y.-F. Liu, and S.-C. Lin, ``Max-min-fairness linear transceiver
  design for full-duplex multiuser systems,'' in \emph{\emph{Proc.} IEEE
  SPAWC}, 2017, pp. 1--5.

\bibitem{kay1998fundamentals}
S.~M. Kay, \emph{\BIBforeignlanguage{eng}{Fundamentals of {Statistical}
  {Signal} {Processing}, {Volume} {II}: {Detection} {Theory}}}, 1st~ed.\hskip
  1em plus 0.5em minus 0.4em\relax Upper Saddle River, NJ: Prentice Hall PTR,
  1998.

\bibitem{Khawar2015}
A.~Khawar, A.~Abdelhadi, and C.~Clancy, ``Target detection performance of
  spectrum sharing {MIMO} radars,'' \emph{IEEE Sensors J.}, vol.~15, no.~9, pp.
  4928--4940, 2015.

\bibitem{liu2024survey}
Y.-F. Liu, T.-H. Chang, M.~Hong, Z.~Wu, A.~M.-C. So, E.~A. Jorswieck, and
  W.~Yu, ``A survey of recent advances in optimization methods for wireless
  communications,'' \emph{IEEE J. Sel. Areas Commun.}, 2024.

\bibitem{KaimingFP2}
K.~Shen and W.~Yu, ``Fractional programming for communication systems—part
  {II}: Uplink scheduling via matching,'' \emph{IEEE Trans. Signal Process.},
  vol.~66, no.~10, pp. 2631--2644, 2018.

\bibitem{GoluVanl96}
G.~H. Golub and C.~F. Van~Loan, \emph{Matrix Computations}, 3rd~ed.\hskip 1em
  plus 0.5em minus 0.4em\relax The Johns Hopkins University Press, 1996.

\bibitem{HouICCC2022}
Y.~Hou, S.~Cai, W.~Xia, J.~Zhang, and S.~Xia, ``Self-interference cancellation
  based on symbol-level precoding for dual-function radar and communication
  systems,'' in \emph{\emph{Proc.} IEEE/CIC Int. Conf. Commun. in China}, 2022,
  pp. 140--145.

\end{thebibliography}

\end{document}